\documentclass[twocolumn,10pt]{revtex4-1}
\usepackage{graphicx,amsmath}

\begin{document}
\title{Generation of ideal thermal light in warm atomic vapor}

\author{J.~Mika, L.~Podhora, L.~Lachman, P.~Ob\v sil, J. Hlou\v sek, M.~Je\v zek, R.~Filip, and~L.~Slodi\v cka}\email{slodicka@optics.upol.cz}

\affiliation{Department of Optics, Palack\' y University, 17. listopadu 1192/12,  771~46 Olomouc, Czech Republic}

\begin{abstract}
We present the experimental generation of light with directly observable close-to ideal thermal statistical properties. The thermal light state is prepared using a spontaneous Raman emission in a warm atomic vapor. The photon number statistics is evaluated by both the measurement of second-order correlation function and by the detailed analysis of the corresponding photon number distribution, which certifies the quality of the Bose-Einstein statistics generated by natural physical mechanism. We further demonstrate the extension of the spectral bandwidth of the generated light to hundreds of MHz domain while keeping the ideal thermal statistics, which suggests a direct applicability of the presented source in a broad range of applications including optical metrology, tests of robustness of quantum communication protocols, or quantum thermodynamics.
\end{abstract}

\maketitle

\section{Introduction}

Generation of light statistics has been of paramount importance for understanding various phenomena in statistical and quantum optics since the presentation of pioneering experiments by R. Hanbury Brown and R. Q. Twiss in 1956~\cite{Han1956a,Han1956b}. The advent of lasers has triggered a prompt application of methods for estimation of light statistics
on studies in broad range of research directions ranging from atomic spectroscopy to optical imaging and metrology, reaching far beyond the areas of pure optical physics and astronomy. The strong promise of application of well controlled quantum systems in quantum communication networks in last two decades led a large part of quantum optics community to focus on the development and characterization of the single photon light sources~\cite{Eis2011,Lod2015}. Together with the further development of field of quantum statistical optics~\cite{mandelWolf}, the generation and control of nonclassical light states stimulated enormous effort on technical improvements in control of various degrees of freedom of light fields at single photon level and their efficient and fast detection, a crucial requirements for direct observability of intrinsically quantum properties~\cite{Bre1997,Lvo2001}. These technical developments have been extensively applied on witnessing of subtle characteristics of nonclassical states of light in experiments mostly represented by observation of good approximations of single-photon states~\cite{Eis2011,Lod2015}. Ideal single-photon states have vanishing variance of the photon number distribution accompanied with characteristic strong antibunching in the measured degree of second-order coherence.

However, a large number of research directions in optical physics has recently rediscovered the strong application potential of light fields with exactly opposite statistical properties, that is, strong photon bunching and super-Poissonian fluctuations of the photon number. From an infinite set of these states, the ideal thermal state corresponding to idealized population of the single light field mode by thermal excitations is of very significant importance. Thermal state has a maximum entropy for the given constant average energy and so it uniquely determines the Bose-Einstein statistics of any mode of thermal radiation. As such, this statistics naturally appears at thermal equilibrium. Ideal thermal radiation can be directly applied in diagnostics of quantum states and processes~\cite{Zam2005,Har2014,Dov2012}, enhancement of nonlinear effects and metrology~\cite{Qu1992,Jec2013,Spa2017}, quantum imaging~\cite{Zha2005,Val2005,Spr2016}, generation of nonclassicality~\cite{Zav2007}, or pioneering tests of quantum thermodynamics~\cite{Vid2016} and quantum key distribution~\cite{Wee2012}.

Although coherence of thermal light field can be also fully described by the classical Maxwell theory of light waves, their precise experimental generation and detection shares number of difficulties typically associated with preparation of single photon states. These include strict requirements on the observed field modeness and photon detection bandwidth, particularly when aiming for direct observation of exact thermal statistics of photons from a source at thermal equilibrium. These difficulties have been recognized by number of experimental teams and led to a broad employment of pseudo-thermal light sources typically based on the scattering of a laser beam from rotating ground-glass disc~\cite{Vid2016,Are1965,Ros1971,Est1971}. Although such approach can emulate thermal light field with very high fidelity, it is technically limited in the achievable spectral bandwidths due to practically achievable disk rotation speeds, average grain spatial size and focal spot of the scattered laser beam to a few MHz range~\cite{Kuu2017}.
However, it is the actual spectral bandwidth which represents a crucial parameter for potential improvement of waste majority of applications of thermal light~\cite{Zam2005,Har2014,Dov2012,Qu1992,Jec2013,Spa2017,Zha2005,Val2005,Spr2016}. On one side it directly relates to the efficiency of various processes and on the other, it allows for increasing their repetition rates. On the other hand, the finite grain size and the corresponding generated speckle pattern can limit the observable bunching as the area from which the observed light is collected approaches its size. This effectively corresponds to the increase in the number of collected light modes~\cite{Vid2016,Kuu2017}.

Observation of a photon bunching corresponds to an increased probability of detecting a photon after a time interval $\tau$ from the previous photon detection compared to a measurement on a stream of independent photons. This conditional probability is proportional to product of intensity and second-order correlation function $g^{(2)}(\tau)$~\cite{mandelWolf}.
The first realization of a light source with directly observable ideal bunching, besides the pseudo-thermal light sources utilizing the bare intensity modulation of coherent light beam, was reported recently in~\cite{Nak2010}. The generation of light was based on the excitation of laser-cooled atomic cloud and the provided theoretical analysis of the second-order coherence quantitatively supported the observed temporal profiles. However, the bare observation of the ideal photon bunching does not certify the observation of the thermal light field mode~\cite{mandelWolf}. At the same time, besides the relatively high technological demands given by the necessity of laser cooling and isolation of atoms from the thermal environment, the source presented in~\cite{Nak2010} imposes strong limits on the easily achievable spectral widths of the emitted radiation due to narrow emission spectra of the bare atomic transitions.

A number of experimental teams have recently attempted to realize a thermal light source with large spectral bandwidth either as a principal resource or as a side result of attempts to engineer a nonlinear interaction with a high degree of mode purity~\cite{Dus2016,Bla2009,Shu2016,Guo2017,Boi2009,Jec2013,Kur2017,Bla2011}. Out of these, the sources utilizing the process of amplified spontaneous emission~\cite{Jec2013,Boi2009,Kur2017,Bla2011} seem to offer technically feasible spectral bandwidth in a few nanometer regime. However, similarly to several characterizations of light statistics coming from blackbody light sources~\cite{Tan2014,Liu2014}, the extreme spectral width practically hinders a direct confirmation of ideal thermal statistics with currently available single-photon detectors. Typical examples of the limit imposed by the detection jitter being much larger than the coherence time of the detected light include attempts with sources of spontaneous parametric down-conversion with generated photon bandwidth at the order of many GHz~\cite{Bla2009}. Several promising techniques have been developed to circumvent the limited detection bandwidths for observation of the photon bunching on short time scales, including the schemes based on the two-photon absorption~\cite{Boi2009}, or nonlinear sum frequency generation~\cite{Fri1985,Abr1986,Qu1992}. Still, even these remarkable technological advancements did not lead to direct unambiguous observation of ideal bunching with large spectral bandwidth.

Here, we report on a direct measurement of the ideal photon bunching and Bose-Einstein photon number probability distribution on light generated by the excitation of warm atomic vapor. The observability of the ideal photon bunching is achieved by the satisfaction of two fundamental experimental requirements, high single modeness of the detected light and sufficiently narrow spectral bandwidth on the order of tens of MHz relative to the timing jitter of the employed single photon detectors of a few hundred picoseconds. These observation is then complemented by directly measurable Bose-Einstein statistics of generated light, which allows the saturation of entropy for given mean photon number. We demonstrate the traceable extension of the presented thermal light source to observable spectral bandwidths at the order of Doppler-broadened atomic transition lines. We provide an experimental guideline for achieving the large bandwidth and outline its limits in the presented experimental platform.

\section{Generation of ideal thermal light}
\label{SecExp}

\subsection{Experimental scheme}

\begin{figure*}[!t]
\centering\includegraphics[width=0.8\linewidth]{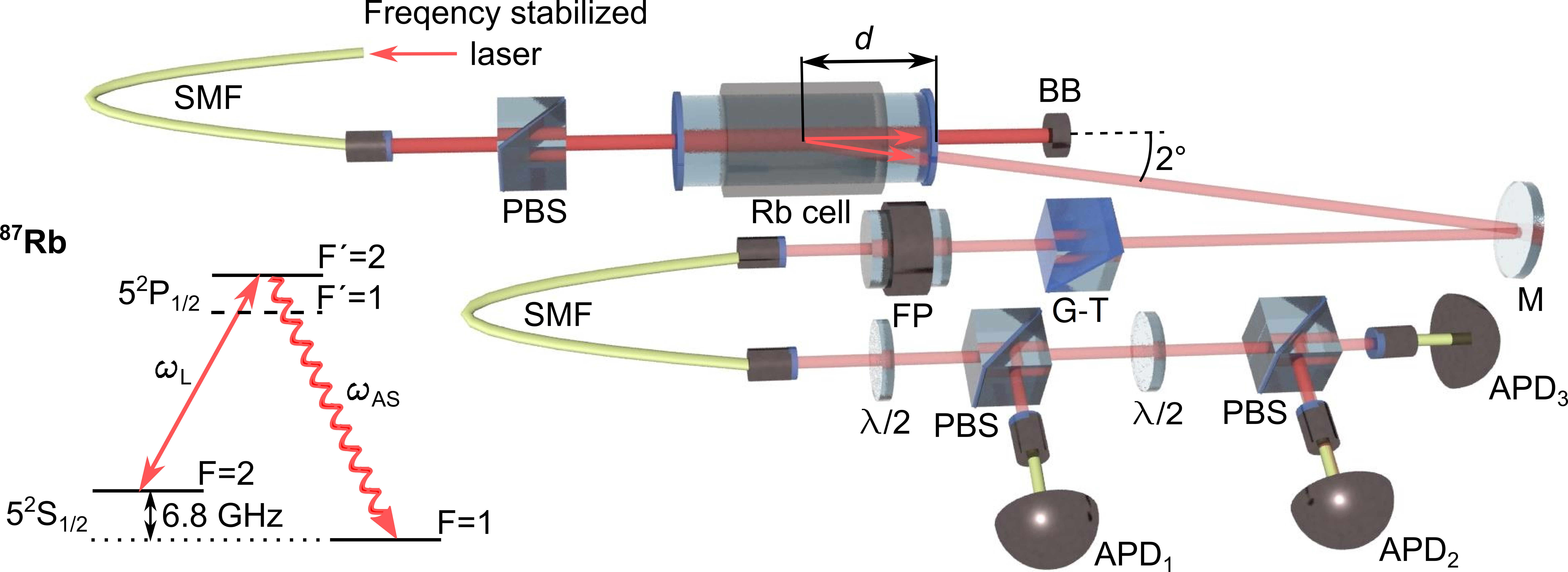}
\caption{The schematic setup of the presented experiment and employed $^{87}$Rb energy level scheme. The polarization and spatial mode of the laser beam emitted from the frequency stabilized laser diode are defined by the input single-mode optical fiber (SMF) and polarization beam splitter (PBS). The laser beam excites an ensemble of warm $^{87}$Rb atoms and the scattered light is observed under small angle of approximately $2^\circ$. The mode of the detected light field is defined precisely using the combination of a Glan-Thompson polarizer (G-T), Fabry-P\'erot resonator (FP) and single mode fiber. The photon statistics of the field is analyzed using an array of single-photon avalanche detectors (APD$_{1-3}$) with balanced detection efficiencies. The balancing is achieved by tuning the $\lambda/2$ polarization half-wave plates in front of each PBS. BB marks the beam block, M is the optical mirror and $d$ is the spatial distance between the observation region and atomic cell output window.} \label{fig:scheme}
\end{figure*}

Our experimental demonstration is based on phenomenologically simple excitation of warm $^{87}$Rb vapor with a laser beam in a single-pass configuration and observation of the scattered light field after precise spatial, polarization and spectral filtering, with the spectral filtering bandwidth surpassing the limit imposed by the finite timing jitters of the employed single photon detectors and the energy splitting of the excited atomic states hyperfine manifolds.
The relevant energy level scheme is depicted in figure~\ref{fig:scheme}. We employ a travelling-wave resonant excitation of warm $^{87}$Rb atoms on D1-line 5S$_{1/2}({\rm F}=2) \leftrightarrow {\rm 5P}_{1/2}({\rm F}'=2)$ transition and frequency selection of the anti-Stokes field generated through Raman transition 5S$_{1/2}({\rm F}=2) \rightarrow {\rm 5P}_{1/2}({\rm F}'=2)\rightarrow {\rm 5S}_{1/2}({\rm F}=1)$ 
with frequency $\nu_{\rm AS}\sim\nu_L+6.8\,{\rm GHz}$. 
Here $\nu_{\rm L}$ is the excitation laser frequency and 
6.8\,GHz corresponds to the splitting of $^{87}$Rb 5S$_{1/2}$ ground state hyperfine manifold.

The excitation laser beam with wavelength of 795~nm is derived from the frequency stabilized laser diode and its spatial mode is set by a single-mode optical fiber (SMF), see figure~\ref{fig:scheme} for the simplified scheme of the experimental arrangement. The polarization of the output Gaussian beam is adjusted to horizontal with respect to the plane defined by the optical table. The laser excites a warm atomic ensemble enclosed in a cylindrical glass cell of 7.5\,cm length and 2.5~cm diameter with antireflection coated windows and with the Gaussian beam radius at the cell position of 440$\pm 10\,\mu$m. The cell is filled with isotopically pure $^{87}$Rb and contains no buffer gas or atomic polarization preserving coatings. The polarization of atomic fluorescence scattered at an angle of $2^\circ$ is filtered to linear polarization perpendicular to the polarization of the excitation beam to suppress the detection of the light coming from the residual laser reflections. We note that the observation of the thermal light statistics was performed for range of observation angles limited by the possibility of sufficient spatial separation from the excitation laser beam for minimal values and by the aperture of the Rb vapour glass cell mount corresponding to about 10\,degrees for maximal achievable scattering angles, without observable effect on the value of the measured bunching. The scattered fluorescence further passes a Fabry-P\'erot resonator with the measured free spectral range FSR$=8854\pm 6$~MHz and the transmission linewidth of $\Delta\nu_{\rm FP} =67\pm 7$~MHz (FWHM). The resonator transmission frequency can be precisely tuned on the scale of its FSR by controlling the temperature of its Zerodur glass spacer. The FSR has been chosen to be larger than the $^{87}$Rb emission spectra which guarantees the selection of a single particular emission frequency mode and, at the same time, the sufficiently narrow transmission linewidth $\Delta\nu$ allows for the selection of at most single radiative transition in the $^{87}$Rb D$_1$-line manifold. The relatively small transmission bandwidth of the employed optical resonator is also crucial for the unambiguous direct detection of the ideal thermal state, as it narrows the emitted Doppler-broadened emission spectra to bandwidths within the limit imposed by finite detection jitters on the order of hundreds of ps of conventional single-photon avalanche detectors operating in the visible spectral range~\cite{Eis2011}. After passing the optical cavity, the atomic fluorescence is coupled to a single-mode fiber which guarantees the spatial purity of the detected field mode. The fluorescence is analyzed in a spatially multiplexed array of three single photon avalanche photodiodes (APDs, Excelitas Technologies, SPCM CD3432H), where the effective individual detection efficiencies have been adjusted to close to equal values using a simple polarization-manipulation scheme comprising the polarization rotation by half-wave plates (HWP) and spatial separation on polarization beam splitters (PBS). The overall photon detection efficiency of 11\,\%. It consists of photon losses on the path from atoms to the output of the single-mode fiber corresponding to about 51\,\%, with the largest contributions from the measured transmissivity of the Fabry-P\'erot filtering resonator, fiber coupling efficiency, and Glan-Thompson polarizer transmission corresponding to 66\,\%, 85\,\%, and 92\,\%, respectively. The losses behind the spatial mode defining single-mode fiber are mostly given by the coupling efficiencies to fiber-coupled single-photon detectors, polarization beam splitter transmissivity and efficiencies of particular single-photon detectors. Balancing of the detection setup results in the detection efficiency of approximately 23\,\% per each detection channel.

The presented photon-detection configuration allows for evaluation of typical statistical characteristics of thermal light manifested in the shape and absolute values of the normalized second-order correlation function $g^{(2)}(\tau)=\langle a^\dagger (t) a^\dagger(t+\tau) a(t+\tau) a(t) \rangle /\langle a^\dagger a\rangle^2$, where $a^\dagger$ and  $a$ are the single-mode creation and annihilation operators, respectively. For that purpose we have used the standard direct measurement approaching  $g^{(2)}(\tau)$ for small mean number of photons~\cite{Gra86}. At the same time, this detection scheme enables the estimation of photon number probability distribution with up to photon numbers limited by the overall measurement length, detection losses and photon generation rate. The detected photon clicks arrival times are recorded using a time tagging device with maximum resolution of 81~ps.

\subsection{Observation of ideal bunching}

\begin{figure}[!t]
\centering\includegraphics[width=0.9\linewidth]{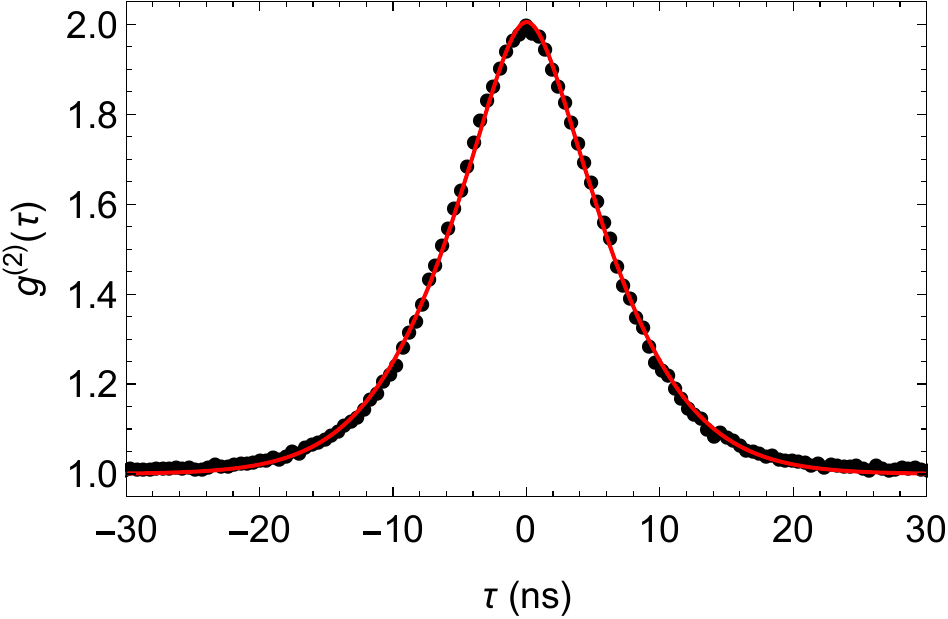}
\caption{The measured second-order correlation function $g^{(2)}(\tau)$. The measured data are represented by points, the solid red line corresponds to the theoretical fit by the simple theoretical model corresponding to the ideal single-mode thermal light. The size of the displayed data points corresponds to the single standard deviation error-bars.} \label{fig:g2}
\end{figure}

The results of the measurement of second-order correlation function for the laser frequency set on the resonance with the $5S_{1/2}(F=2) \leftrightarrow 5P_{1/2}(F'=2)$ transition and cavity transmission peak set to the energy difference between 5S$_{1/2}({\rm F}=1) \leftrightarrow {\rm 5P}_{1/2}({\rm F}'=2)$ levels are shown in figure~\ref{fig:g2}. The excitation laser beam power has been set to $36$~mW for which we have obtained the resulting average count rate on the three detectors $(37\pm 4)\times10^4 $~counts/s.  The spatial overlap of excitation and observation modes has been positioned close to the output optical window of the atomic vapor cell which guarantees a small absorbtion of the emitted resonant light. The vapor cell is set to a temperature of $61^\circ$C which corresponds to the optical thickness of $0.78\pm 0.05$ for the resonant excitation of 5S$_{1/2}({\rm F}=2) \leftrightarrow {\rm 5P}_{1/2}({\rm F}'=2)$ transition. The displayed $g^{(2)}(\tau)$ corresponds to the average of the measurement on all three possible two-detector combinations with the overall measurement time of 173~minutes and the coincidence-detection window set to 648~ps. The evaluated peak value of $g^{2}(0)=2.00\pm 0.01$ from the data presented in the figure~\ref{fig:g2} suggests the generation of ideal thermal light field. Since we aim to evaluate the light source, the presented $g^2(\tau)$ has been corrected for the residual intrinsic detector dark counts of $(350\pm 20, 180\pm 13, 265 \pm 18)$ counts/s measured on the employed APD$_1$, APD$_2$ and APD$_3$, respectively. The raw value of $g^{(2)}(0)$ without correction on detector dark counts is $1.98\pm 0.01$. We note, that these results have been achieved without any particular experimental optimization procedure and the generated statistics corresponds to a natural output of the presented scheme in a broad range of spatial alignment and excitation frequency settings. The red line in figure~\ref{fig:g2} corresponds to a theoretical fit using the normalized second-order correlation function for ideal single-mode thermal light $g^{(2)}(\tau)=1+|g^{(1)}(\tau)|^2$~\cite{mandelWolf}, where the first-order correlation function $g^{(1)}(\tau)$ is modelled by assuming the Doppler broadened emission on the 5P$_{1/2}({\rm F}'=2)\rightarrow {\rm 5S}_{1/2}({\rm F}=1)$ transition followed by the spectral filtration by the Fabry-P\'erot resonator with independently measured transmission bandwidth. The frequency bandwidth $\Delta\nu =65\pm 3$~MHz (FWHM) of detected light has been estimated by numerically evaluating the temporal width of the measured $g^{(2)}(\tau)$ function~\cite{Dus2016}, in good agreement with the independently measured transmission width of the optical filtering cavity $\Delta\nu_{\rm FP} =67\pm 7$~MHz.

\subsection{Measurement of the Bose-Einstein probability distribution}

\begin{figure}[!t]
\centering\includegraphics[width=1.\linewidth]{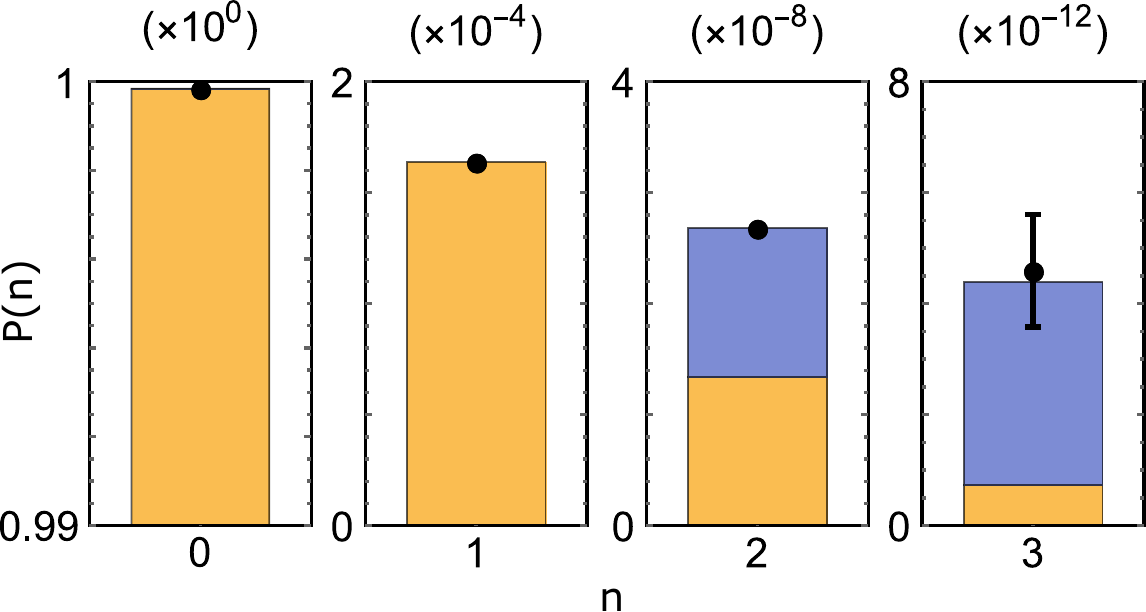}
\caption{The estimated photon number probability distribution (black points) with the blue and yellow bars corresponding to the theoretical photon number distributions for ideal thermal and coherent light fields with the same mean photon number, respectively, where the visible blue part shows the difference between these statistics. Error bars correspond to estimated single standard deviations, which are for the first three bars (n={0,1,2}) smaller than the data point.}
\label{fig:pns}
\end{figure}

To unambiguously clarify our conclusions about the detection of the ideal thermal state, we measure the photo-count statistics from which the photon number probability distribution $P(n)$ can be estimated. As the probability of the four-coincidence events for the observed thermal light count-rates within the chosen temporal mode width of 648~ps would be practically negligible even on the measurement time scales of days, it suffices to employ a three-detector scheme. The photon number probability distribution is estimated from the measured rates of singles, two-fold and three-fold coincidences using both the direct inversion~\cite{Spe2012} and the maximum-likelihood estimation~\cite{parisRehacek}. The distributions resulting from the two methods differ only negligibly and within the largest error bar at $P(3)$ which has negligible effect on evaluated metrics, and we thus use and display only the distribution based on maximum likelihood algorithm, see figure~\ref{fig:pns}.

The estimated $P(n)$ corresponds to the ideal single-mode thermal light within statistical measurement errors, which certifies the unambiguous preparation of the ideal thermal light statistics. To illustrate the significance of the estimation of higher photon numbers in the presence of large attenuation, we show also the statistics of the ideal coherent state with the same mean photon number $\langle n\rangle=(1.64\pm 0.01)\times10^{-4}$ for comparison. The evaluated $g^{(2)}(0)=2.00\pm0.01$  from full measured photon statistics corresponds to Mandel-Q parameter $Q=(1.64\pm 0.01)\times10^{-4}$, which is in good agreement with the theoretically expected value $Q_{\rm therm}=1.64\times10^{-4}$ for the ideal thermal state with given mean photon number. To characterize the quality of observed Bose-Einstein statistics, we employ the fundamental definition of the thermal light: it is light with maximum Shannon entropy $H=-\sum_n p(n) \log p(n)$ for the given mean number of photons. For the presented data $H=(181175\pm 4)\times10^{-8}$ in agreement with the theoretically expected $H_{\rm theory}=181175\times10^{-8}$. The actual distance between the ideal Bose-Einstein distribution and the measured one can be quantified by evaluation of the relative entropy $H(P(n)|P_{\rm ref}(n))$, where $P_{\rm ref}(n)$ corresponds to the reference probability distribution~\cite{Kul1951}. Advantageously, this distance is operational, it has a thermodynamical interpretation in terms of work necessary to be done to reach the ideal thermal state from the measured one~\cite{Esp2011}. The evaluated relative Shannon entropies for the measured statistics with respect to ideal thermal and coherent state with the same mean photon numbers are $H(P(n)|P_{\rm therm}(n))=(2\pm 2) \times 10^{-14}$ and $H(P(n)|P_{\rm coh}(n))=(516\pm 1)\times 10^{-11}$, respectively. They further certify the relative thermodynamical closeness of the generated state to the ideal thermal state compared to a coherent state with the same energy. The uncertainties of the evaluated entropies have been estimated using the Monte Carlo routine from the uncertainties of the measured numbers of photon clicks. These results qualify the realized single-mode thermal light source for experiments in quantum thermodynamics~\cite{Vid2016}, where the ideal thermal light statistics is required to reliably
emulate thermal equilibrium quantum states corresponding to basic energy resources for thermodynamical tests at single photon level~\cite{Hlo2017}.

\section{Thermal light with tunable bandwidth}

The generation of thermal light using atomic vapors with high mean velocity thermal motional distribution corresponding to large Doppler broadening can naturally benefit from it and can be directly extended to frequency bandwidths on the order of several hundreds of MHz. The large spectral bandwidth can enhance the efficiency of interaction of the generated thermal light with target atomic ensembles and allow higher repetition rates of its potential applications. We experimentally characterize such extension by measurement of the statistics of light emitted from warm Rb vapor by tuning several crucial parameters which directly influence the spectral and temporal width of the observed fluorescence. To allow the observation of narrow temporal wave packets, we replace the optical frequency filter in our setup by the Fabry-P\'erot cavity with the free spectral range of about FSR $=30$\,GHz and measured transmission linewidth of $\Delta\nu_{\rm FP} =818\pm 3$~MHz (FWHM). The chosen $\Delta\nu_{\rm FP}$ is sufficiently narrow so that the contribution of emission from the F $=1$ state of the 5P$_{1/2}$ manifold to the detected light is negligible and, at the same time, it allows for capturing of almost whole spectral width of Doppler-broadened emission for the considered Doppler broadening of about $\sigma_D=260$\,MHz. On the other hand, the large bandwidth of emitted photons will technically compromise the direct observability of ideal thermal correlations due to the finite timing resolution of single-photon detectors. However, the precise knowledge of the exact time response of particular employed detectors allows recovering the true $g^{(2)}(0)$ values with high confidence.

\begin{figure}[!t]
\centering\includegraphics[width=0.9\linewidth]{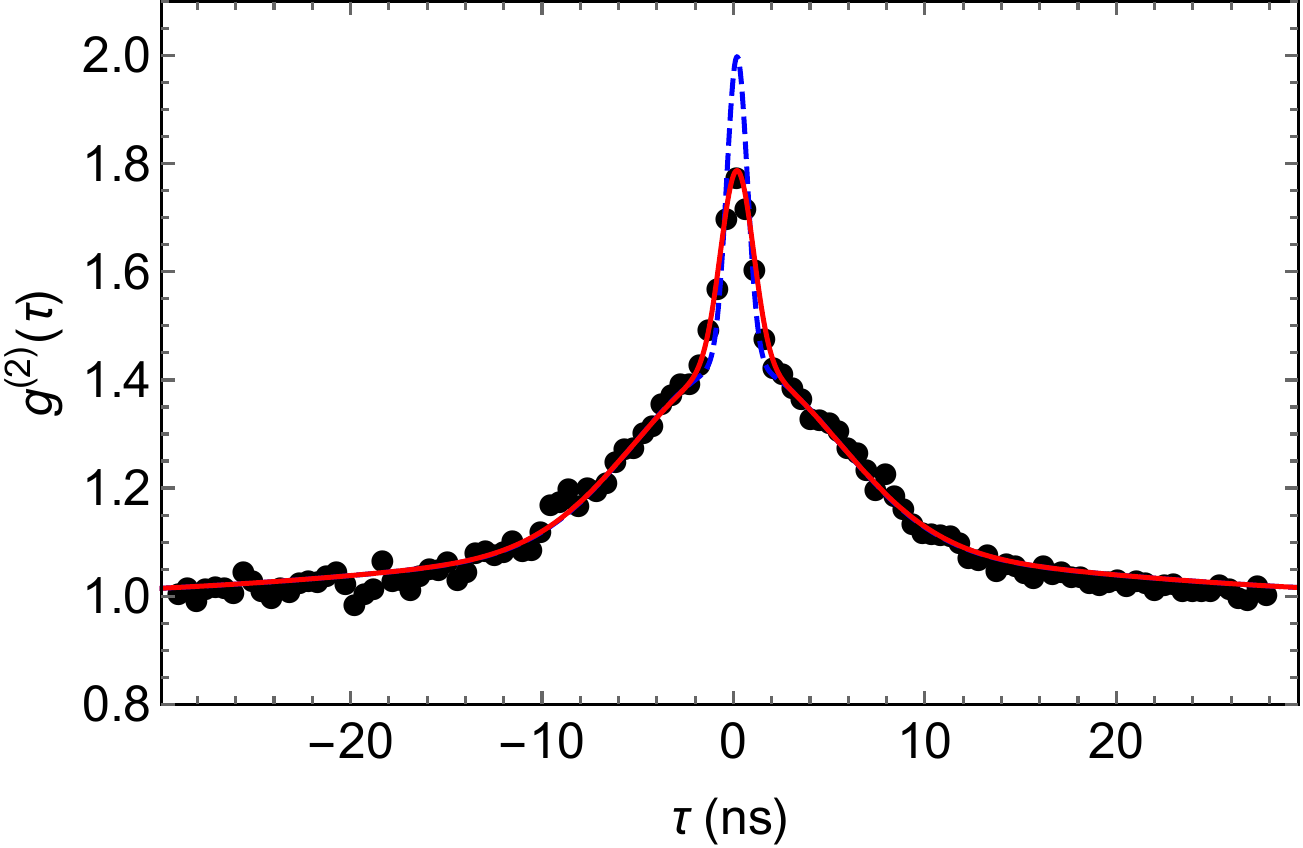}
\caption{The measured $g^{(2)}(\tau)$ with the broadband optical filtering fitted by the coherent superposition of Gaussian functions corresponding to single and multi-photon scattering (red solid line), see text for details. The blue dashed curve corresponds to the $g^{(2)}(\tau)$ obtained by the correction of the measured function on exactly measured timing uncertainties of employed single-photon detectors. The estimated error bars correspond to the size of displayed data points.} \label{fig:g2broad}
\end{figure}

Figure~\ref{fig:g2broad} shows a typical measured second-order correlation function $g^{(2)}(\tau)$ with the broadband Fabry-P\'erot optical filter. The excitation laser power has been set to $P=39$\,mW and the atomic cell temperature is corresponding to effective optical depth ${\rm OD} = 1.19\pm 0.08$. The observed temporal shape corresponds to two scattering processes, single-atom scattering and multi-atom scattering of detected photons, which give rise to two characteristic extents of reconstructed spectral widths~\cite{Dus2016,Car2015}. The presented data have been fitted using a simple theoretical model comprising two Gaussian spectra and the optical cavity filter with Lorentzian profile resulting in the estimated ratio of single to multi-photon scattering of $0.15$. Importantly, the maximal value of $g^{(2)}(0)=1.71\pm 0.03$ still corresponds to the ideal thermal state if we consider the timing jitter of employed detectors, whose effect can be unambiguously accounted for using the deconvolution of the measured $g^{(2)}(\tau)$ with their exact response function. The timing jitters have been precisely characterized in an independent measurement using fast mode-locked pulsed laser with pulse lengths of about 200\,fs attenuated to single photon level and precise recording of the photon detection time. The evaluated timing uncertainties of $\sigma_1=358$\,ps, $\sigma_2=551$\,ps, $\sigma_3=365$\,ps for APD$_1$ to APD$_3$, respectively, result in the estimated $g^{(2)}(0)=2.00\pm 0.04$, a clear signature of preserving the photon statistics of ideal thermal light. The frequency bandwidth of the detected light field has been estimated from the observed time spread in $g^{(2)}(\tau)$, which gives the temporal standard deviation $\sigma_{\tau}=13\pm 2$\,ns corresponding to the frequency width $\sigma_{\nu}=110\pm 20$\,MHz. The $\sigma_{\tau}$ and $\sigma_{\nu}$ were evaluated following the calculation presented in Dussaux et al.~\cite{Dus2016}. The detailed experimental studies of the dependence of generated thermal light bandwidth and detected photon rate on parameters which determine the interaction of light with atomic vapor can be found in appendix.

\section{Conclusion and outlook}

We have demonstrated the generation of ideal thermal light field from a warm atomic vapor by measurement of corresponding photon-number statistical properties. The presented results substantially enhance the observable spectral bandwidths of light fields with the ideal photon bunching and, at the same time, the employed evaluation methods for the first time unambiguously prove the thermal light photon statistics. The observed $g^{(2)}(0)=2.00\pm 0.01$ together with the reconstructed ideal Bose-Einstein statistical distribution without any correction on the finite detector time response from a natural light source represent the achievement of a steadily pursued goal since pioneering experiments on detecting the statistical correlations of light from low-pressure glow discharge lamps~\cite{Mar1964,Mor1966}. We conclusively confirmed that Shannon entropy of the measured probability distribution corresponds exactly to theoretically predicted one for the measured mean number of photons. Besides its strong fundamental interest, we have shown that the thermal light generated from warm atomic vapor can benefit from the large Doppler-broadened spectral width of emitted photons, which is directly proportional to the atomic temperature. In addition, the bandwidth scalability to $\sigma_\nu=270$~MHz goes in hand with the extreme technological simplicity of the scheme based on the bare single-pass Raman excitation of a warm atomic vapor which can be easily adapted for vast majority of atomic species used commonly in quantum optics experiments.

Together with the detected photon rates of $3.4\times 10^5$\,counts/s, these results promise enhancement of a large number of applications in classical and quantum optics~\cite{Zha2005,Val2005,Zam2005,Spa2017,Jec2013,Har2014,Spr2016,Qu1992,Dov2012}. The presented source possesses several important advantages compared to conventional thermal light sources based on amplified spontaneous emission~\cite{Jec2013,Boi2009,Kur2017,Bla2011} or parametric down conversion~\cite{Dov2012,Bla2009,Guo2017}. While sharing their technical simplicity, it allows for direct detection of ideal bunching with conventional visible-range single-photon detectors, a crucial property for its feasible benchmarking. At the same time, its spectral properties allow for a direct utilization in interactions with target atomic ensembles, in which prospective tunable nonlinearities and efficient and large bandwidth light storage have been demonstrated~\cite{Shu2016,Kai2009,Wal2018,Fir2018,Luk2005,Buc2011}. We foresee employment of the developed source for studies of optical nonlinearities in quantum regime and of effects of photon statistics on the efficiency of the nonlinear processes~\cite{Spa2017,Kai2009}. We aim for the utilization of the presented thermal light for investigation of noise properties of single-photon level optical memories based on warm atomic ensembles~\cite{Wal2018}. The degree of detected photon bunching and fidelity of the retrieved photon number distribution with the ideal Bose-Einstein statistics will allow to reveal the amount and statistical structure of memory-added noise, as the ideal photon bunching observability critically depends on the light modeness~\cite{Dov2012}. The generated thermal states can be also readily employed for the production of single-photon-added thermal states, which were suggested as a promising resource for testing criteria of nonclassicality~\cite{Zav2007}. They also have a potential to be useful for reconstruction of photon-number distributions~\cite{Har2014}, where precise generation and measurement of thermal light statistics can play crucial role.

\section*{Acknowledgments}
We acknowledge the financial support of the Czech Science Foundation by grant No.~GA14-36681G and Palack\'y University
IGA-PrF-2017-008.

\section*{Appendix: Tunability of the observed spectral bandwidth}

\setcounter{figure}{0}
\renewcommand{\thefigure}{A\arabic{figure}}


The overall temporal length of the generated thermal light is substantially broadened by the contribution of the single-atom or few-atom scattering processes. As demonstrated in references~\cite{Dus2016,Hol1947}, the contribution of the single atom scattering process to the observed $g^{(2)}(\tau)$ can be suppressed by maximization of the effective optical thickness, which can be achieved by resonant excitation and high atomic density. We have further examined the dependence of the overall generated field frequency bandwidth and possibility of keeping the observability of the ideal photon bunching on several crucial parameters determining the interaction of light with atomic vapor and relative contribution of the single to multi-atom scattering. These include the effective optical thickness of the atomic sample, excitation laser power, laser detuning from the 5S$_{1/2}({\rm F}=2) \leftrightarrow {\rm 5P}_{1/2}({\rm F}'=2)$ atomic resonance and the spatial position of the observation region with respect to the output optical window.

\begin{figure}[!t]
\centering\includegraphics[width=1.0\linewidth]{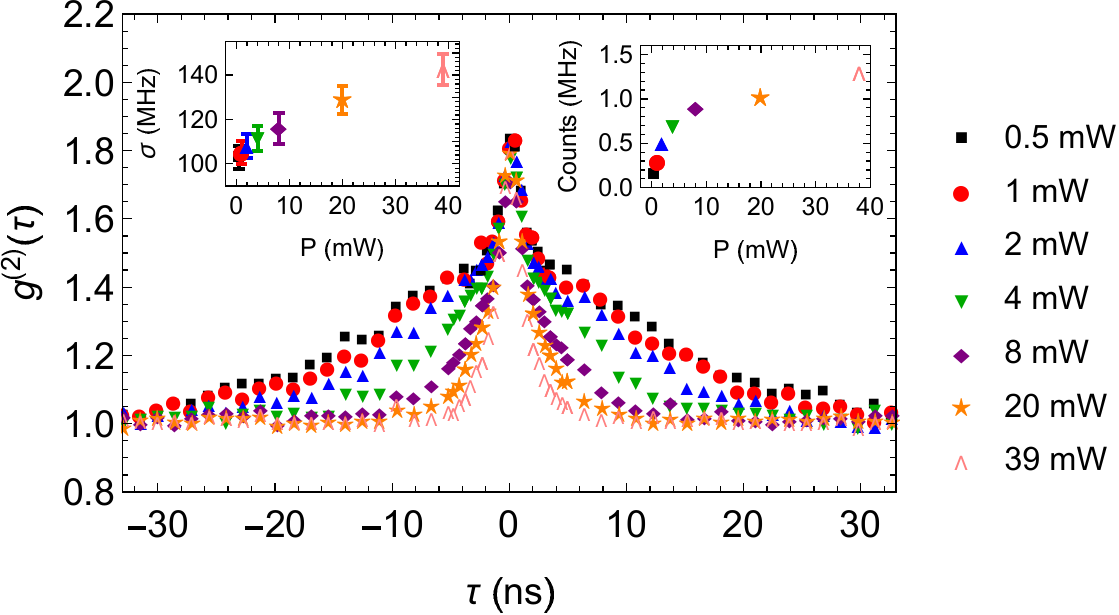}
\caption{The evaluated $g^{(2)}(\tau)$ functions measured for various powers of excitation laser beam and position of the excitation and observation spatial modes overlap at approximately d=0\,cm, that is at the output window of the vapor cell. The two insets show the measured average count rates (right) and estimated thermal light frequency bandwidths (left). The estimated error bars correspond to a single standard deviation and are, where not directly illustrated, roughly the same size as plot markers.} \label{fig:power1}
\end{figure}

\begin{figure}[!t]
\centering\includegraphics[width=1.0\linewidth]{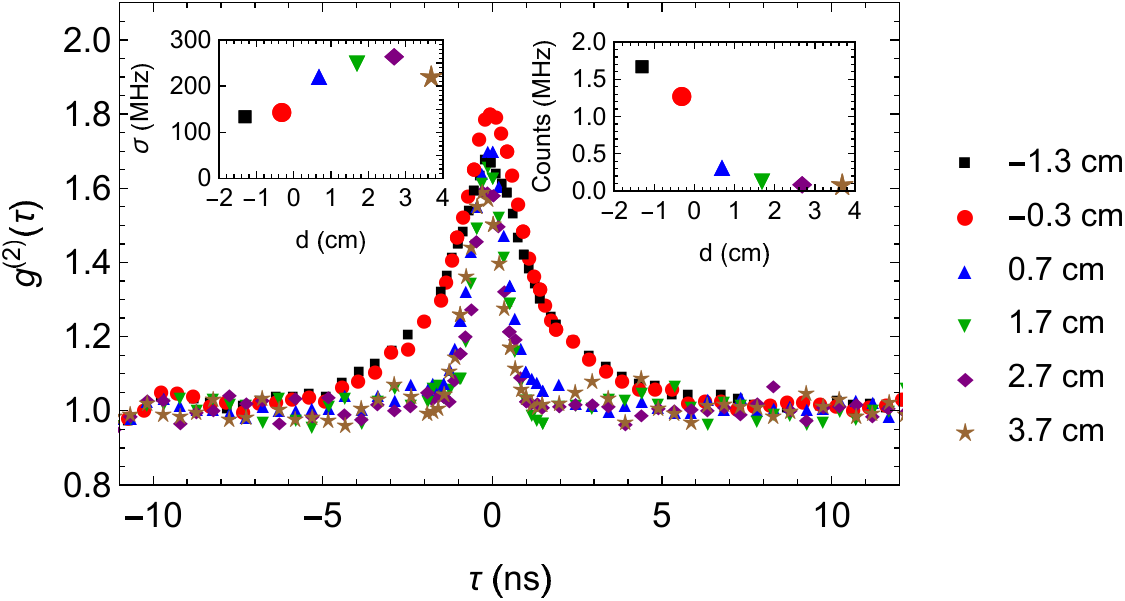}
\caption{The measured $g^{(2)}(\tau)$ of scattered light for various distances $d$ of the rubidium cell output window from the center of the observation region defined by the intersection of the excitation and observation beam modes, see text and figure~\ref{fig:scheme} for more details. The insets show the corresponding estimated light field spectral width (left) and detected count rate (right). The statistically estimated error bars correspond to one standard deviation, where indicated, otherwise they are below the extent of displayed data points.} \label{fig:displacement}
\end{figure}

The following measurements are realized for optical thickness of $1.19\pm 0.08$ corresponding to maximal easily achievable temperature $70^\circ$C in our setup and resonant excitation on the 5S$_{1/2}({\rm F}=2) \leftrightarrow {\rm 5P}_{1/2}({\rm F}'=2)$ transition. The observed power dependence of the $g^{(2)}(\tau)$ shown in the figure~\ref{fig:power1} suggests the convenience of increasing power for both the absolute observable spectral bandwidth and the detected photon count rate with the maximum values of $\sigma_{\nu}=142\pm7$~MHz and $1.29\times10^6$\,counts/s, respectively, corresponding to the maximal excitation laser power of 39\,mW. We note that for all probed input laser powers, the observed values of $g^{(2)}(0)$ after correction for detector timing jitters lead to ideal thermal state value with accuracy within estimated errors one standard deviation.

We have also examined the possibility of further suppression of the single-atom scattering contribution by changing other measuring the $g^{(2)}(\tau)$ as a function of the spatial distance $d$ defined as a distance between the cell output optical window and the observation region given by the intersection of excitation and observation spatial modes, see Figure~\ref{fig:scheme} for the spatial configuration. We choose $d$ to be positive for the observation regions inside the atomic cell. The suppression of the multi-atom scattering contribution for positions of the excitation-observation region close to the cell output window is caused by the decreased number of atoms in the path of the scattered photons propagating towards the detection setup. The observed dependence presented in Fig.~\ref{fig:displacement} for $39$~mW excitation laser power suggests that positioning the interaction region to $d\geq 0.7$\,cm inside the atomic cell almost fully suppresses the observed single scattering contribution and leads to thermal light spectral bandwidths of up to $\sigma_{\nu}=270\pm 20$\,MHz. However, the increased length of the atomic medium strongly suppresses the overall generated photon rate resulting in $3.4\times 10^5$\,counts/s corresponding to $\langle n\rangle=(1.6\pm0.1)\times10^{-4}$. The scaling of the detected photon rate and spectral bandwidth can be seen in insets of the Fig.~\ref{fig:displacement}. The considered detected photon rate here corresponds to the sum of rates from all three detectors. The observable steep dependence of both the overall detected count rate and contribution of the single-atom scattering around $d\approx 0$ corresponds well to the spatial position of the interaction region at the cell window. To the best of our knowledge, the presented estimated spectral widths of the thermal light fields are by at least two orders of magnitude better than any previously reported values achieved with pseudo thermal light sources~\cite{Kuu2017} and, at the same time, the measured $g^{(2)}(0)$ values corrected for the detector timing jitter suggest the generation of the ideal single-mode thermal light.


\end{document}